\documentclass[pra, twocolumn, showpacs]{revtex4}

\usepackage{verbatim}  
\usepackage{textcomp}
\usepackage{amsmath}
\usepackage{amssymb}
\usepackage{amsfonts}
\usepackage{mathrsfs}
\usepackage{graphicx}
\usepackage{dcolumn}
\usepackage{bm}

\begin{document}

\preprint{APS/123-QED}

\title{Rubidium Pump-Probe Spectroscopy: Comparison between ab-initio Theory and Experiment}

\author{M. Himsworth and T. Freegarde}

\affiliation{School of Physics \& Astronomy, University of Southampton, SO17 1BJ, UK.}
\email{m.himsworth@soton.ac.uk}
\begin{abstract}
We present a simple, analytic model for pump-probe spectroscopy in dilute atomic gases. Our model treats multilevel atoms, takes several broadening mechanisms into account and, with no free parameters, shows excellent agreement with experimentally observed spectra.
\end{abstract}
\maketitle
\section{Introduction}\label{intro}
There has been much interest recently in the effect of the propagation of a laser beam through dilute vapors of alkali atoms \cite{Lindvall, Zigdon, Siddons, Zigdon2, Smith} due to their application in laser cooling \cite{Chu, Migdall}, chip-scale atomic clocks and magnetometers \cite{Gerginov, Shah}, laser stabilization \cite{Tsuchida, Bjorklund, Petelski}, and electromagnetically induced transparency \cite{Rapol}. In order to understand the spectra, several models have been proposed which accurately predict the Doppler-broadened and sub-Doppler absorption spectra of a dilute vapor with weak and strong probe beams \cite{Maguire, Lindvall, Borde, Pappas, Nakayama, Haroche}. The common approach of numerically solving the semi--classical density matrix equations, however, requires intensive computation \cite{Maguire}. This may be unnecessarily complicated for several applications in which coherence effects are negligible and beam powers relatively low, such as studies of optical pumping \cite{Smith}, transit- \cite{Lindvall, Smith, Nakayama, Borde} and power-broadening and number density measurements. In this paper we show that a simple ab-initio model based upon rate equations can predict both Doppler-broadened and sub-Doppler spectra with high accuracy for a large range of beam powers and widths.\par
Our model has been developed primarily to focus upon dilute vapors of alkali atoms, specifically rubidium, but the results should be generally applicable to any dilute vapor. We use the Einstein rate equations to calculate the steady state population densities and therefore spectra are assumed to be measured on timescales greater than any coherence effects \cite{Haroche}. The model also assumes collimated, pump and probe beams, whose spectral widths are less than the spontaneous decay rates of the atoms.\par
This paper begins with a short review of the theory behind saturated absorption spectroscopy; the next section covers the effects of optical pumping on absorption lineshapes with multilevel atoms, which accounts for the majority of observed features. We then compare the result of this simple model with experimental pump--probe spectra of rubidium and show that the fit is accurate, even without any free parameters.
\section{Pump-Probe Theory}\label{pptheory}
The absorption of a weak probe beam propagating in the $z$ direction through a dilute gas is characterized by the Beer--Lambert relation \cite{Foot},
\begin{equation}
\frac{\text{d}I(\omega)}{\text{d}z}=-N_{V}\sigma(\omega)I_{0}\label{eqn1}.
\end{equation}
where $N_{V}$ is the number density of atoms, $I_{0}$ is the incident probe intensity and $\sigma(\omega)$ is the absorption cross section for the electric dipole transition $i\to k$, in which $i$ is the ground state and $k$ is the excited state, and the difference in state energies $E_{k}-E_{i}=\hbar\,\omega_{ik}$: 
\begin{equation}
\sigma(\omega)=\frac{\hbar\,\omega_{ik}}{c}\left(B_{ik}n_{i}-B_{ki}n_{k}\right)L(\omega)\text{d}v\label{eqn2}
\end{equation} 
where $n_{i}$ and $n_{k}$ are the \emph{fractional} populations of the ground and excited states, $B_{ik}$ and $B_{ki}$ are the Einstein coefficients for the absorption and stimulated emission of a photon, respectively, and
\begin{equation}
L(\omega)=\frac{\gamma/2\pi}{(\omega-\omega_{ik})^{2}+\frac{\gamma^{2}}{4}}
\end{equation}
is the normalized Lorentzian function which characterizes the atom's response to the incident field, where $\omega$ is the frequency of the laser and $\gamma=1/\tau$ is the natural linewidth of the transition which has the excited state lifetime $\tau$.

In common with many texts, we set the excitation rate to be proportional to the energy density in each spectral mode $\rho(\omega)$ for a broadband light source. For high resolution spectroscopy we require the measuring tool---the laser---to have a finer resolution than the subject under investigation and therefore the spectral linewidth of the laser must be $\Delta\omega<1/\tau$. Hence the spectral energy density of a nearly monochromatic beam with an electric field amplitude $\mathcal{E}$ may be written as \cite{Corney}
\begin{equation}
\rho(\omega)=\frac{1}{2}\epsilon\,\mathcal{E}^{2}L(\omega)=\frac{I}{c}L(\omega),\label{eqn5}
\end{equation}
which has already been assumed in Equation \ref{eqn2}. For states with degeneracy $g_{i,k}$, the Einstein coefficients are
\begin{equation}
B_{ik}=\frac{\pi e^{2}}{\epsilon_{0}\hbar^{2}g_{i}}\sum_{m_{F}}\vert \mu_{ik} \vert^{2}=\frac{g_{k}}{g_{i}}B_{ki},\label{eqn6}
\end{equation}
where $e$ is the electron charge, $\epsilon_{0}$ is the vacuum permittivity and $\vert\mu_{ik}\vert=\vert\mu_{ki}\vert=C_{ik}\vert\langle J_{k}\vert\vert\vec{r}\vert\vert J_{i} \rangle\vert$ are the dipole matrix elements for electric dipole transitions between states with spin-orbit angular momentum $J_{i,k}$, in which 
\begin{equation}
\vert\langle J_{k}\vert\vert\vec{r}\vert\vert J_{i} \rangle\vert=\mu_{0}=\sqrt{\frac{2J_{k}+1}{2J_{i}+1}}\sqrt{\frac{3\pi\epsilon_{0}\hbar c^{3}}{\tau\omega_{ik}^{3}}}\label{eqn7}
\end{equation}
is the reduced dipole moment \cite{Loudon, Siddons} and $C_{ik}$ are the Clebsch-Gordan coefficients for the $m_{F}$ sublevels which define the relative strength for each transition. For unpolarized beams with no external magnetic fields (no quantization axis) the dipole matrix elements are;
\begin{equation}
\sum_{m_{F}}\vert\mu_{ik}\vert^{2}=\mu_{0}^{2}\sum_{m_{F}}C_{ik}^{2}\label{eqn8}.
\end{equation}
Values of $\sum_{m_{F}}C_{ik}^{2}$ can be obtained from tabulated results \cite{Metcalf} or via a lengthy calculation \cite{Edmonds}.\ par
It is common in many texts to include a prefactor of $1/3$ in Equation \ref{eqn6} to average over the 3 possible orientations of the atom with no quantization axis \cite{Demtroder, Letokhov, Foot}. For degenerate states, the summation over all $m_{F}$ sublevels yields the same value for all polarizations and therefore in Equation \ref{eqn5} we assume linear polarization and avoid the $1/3$ prefactor.

The lineshapes of the resonances will be inhomogeneously broadened by the motion of the atoms. For a dilute vapor, the atoms have a mean free path greater than the dimensions of the cell, therefore collisions are negligible and the spread of velocities will follow a Maxwell-Boltzmann distribution with a mean temperature, $T$, equal to that of the cell walls. We therefore have the distribution of atoms with velocity $v$,
\begin{equation}
f_{D}(v)=\frac{1}{u\sqrt{\pi}}\exp\left({\frac{-v^{2}}{u^{2}}}\right).\label{eqn10}
\end{equation}
and mean speed,
\begin{equation}
u=\sqrt{\frac{2k_{B}T}{M}},\label{eqn11}
\end{equation}
where $M$ is the atomic mass and $k_{B}$ is the Boltzmann constant. The change in intensity for a weak probe beam through a dilute vapor is therefore
\begin{equation}
\frac{\text{d}I(\omega)}{\text{d}z}=-\frac{\hbar\omega_{ik}\gamma}{2\pi c}N_{V}I_{0}B_{ik}\int_{-\infty}^{+\infty}\frac{f_{D}(v)(n_{i}-\frac{g_{k}}{g_{i}}n_{k})}{(\omega-\omega_{ik}-kv)^{2}+\frac{\gamma^{2}}{4}}\text{d}v,\label{eqn12}
\end{equation}  
where $kv=(\omega/c)v$ is the Doppler shift which has been included into the Lorentzian lineshape
\begin{equation}
L(\omega)=\frac{\gamma/2\pi}{(\omega-\omega_{ik}-kv)^{2}+\frac{\gamma^{2}}{4}}
\end{equation}
and similarly with $\rho(\omega, v)$. The integrand has a Voigt lineshape which cannot be solved analytically but may be calculated from tabulated values \cite{Siddons}. 
\subsection{Saturation}\label{saturation}
The common experimental arrangement is for a strong pump beam and weak probe beam, derived from the same laser and hence equal in frequency, to propagate in opposite directions through the atomic sample. The pump beam affects the population difference, $n_{i}-\frac{g_{k}}{g_{i}}n_{k}$, and thus reduces the probe absorption near resonance. Only atoms which travel nearly perpendicular to both beams, and thus have a zero velocity component along the beam direction, will be pumped and probed simultaneously, and therefore narrow 'Lamb dips' form in the Doppler-broadened profile at resonance \cite{Letokhov}.

 Using the Einstein rate equations for a closed, degenerate two level system ($n_{i}+n_{k}=1$, $g_{k}=g_{i}$), the steady state population difference is
\begin{eqnarray}
n_{i}-n_{k}&=&\left(1+\frac{2B_{ki}L(\omega, -v)I_{P}}{c\gamma}\right)^{-1}\label{eqn13}\\
&=&\left(1+\frac{I_{P}}{I_{ik}}\frac{\pi\gamma}{2}L(\omega, -v)\right)^{-1},\label{eqn14}
\end{eqnarray} 
where $I_{P}$ is the pump beam intensity, and
\begin{equation}
I_{ik}=\frac{\pi c \gamma^{2}}{4B_{ki}}=\frac{\hbar^{2}\epsilon_{0}c\gamma^{2}}{4q^{2}\vert \mu_{ik}\vert^{2}}\label{eqn15}
\end{equation}
is the saturation intensity at which $n_{k}=1/4$ on resonance and the negative sign for the velocity in Equation \ref{eqn13} reflects the counter-propagating geometry. As the pump intensity increases, the population difference tends to zero, the atomic sample becomes transparent to further excitation, and thus the probe beam shows decreased absorption.

This model is commonly presented in text books \cite{Foot, Demtroder} and is known as \emph{saturated absorption spectroscopy}. Unfortunately it fails to model realistic systems in two important ways. Firstly, the system is not closed as atoms may enter or leave the beams; this limits the interaction time and re-thermalizes the population difference. Secondly, closed two-level systems are very rare and the atoms will generally have several ground and excited states into which they may be optically pumped. If these additional states are separated by frequencies greater than the laser linewidth, then once these states are populated the atom will become transparent to the laser radiation. This process can occur even at low pump intensities and therefore is a significant factor affecting the depth of the sub-Doppler features \cite{Smith}. 
\subsection{Optical Pumping}\label{optpump}

\begin{figure}[t]
\centering
\includegraphics[width=5cm]{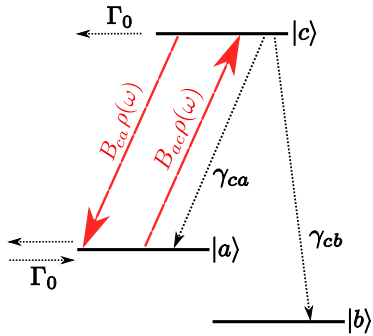}
\caption{\emph{(colour online) Simple 3 level system with optical pumping on $\vert a\rangle\to\vert c\rangle$ transition. The steady state populations in each state depend upon the decay fractions $\gamma_{ik}$ and the populations of atoms entering and leaving the beam at rate $\Gamma_{0}$.}\label{3level}}
\end{figure}

The simplest multilevel system involves a single excited state $\vert c \rangle$ which may decay into two ground states $\vert a\rangle$ and $\vert b\rangle$, as shown in Figure \ref{3level}. The excited state decays at a total rate $\gamma_{c}$ but the fractional rates into each ground state (branching ratio) depend upon the ratio of Clebsch-Gordan coefficients, $\gamma_{ca}$ and $\gamma_{cb}$, where
\begin{eqnarray}
\gamma_{ca}&=&\frac{C_{ac}^{2}}{C_{ac}^{2}+C_{bc}^{2}}\gamma_{c}\\
\gamma_{cb}&=&\frac{C_{bc}^{2}}{C_{ac}^{2}+C_{bc}^{2}}\gamma_{c}.
\end{eqnarray}

Atoms drift into and out of the beam at a rate $\Gamma_{0}$, entering with the equilibrium population distribution and leaving with the optically--pumped distribution. This is discussed further in Section \ref{transit}.

We consider the absorption from state $\vert a\rangle$, so that any population excited from $\vert b\rangle$ is negligible. The rates of change of the state populations are therefore
\begin{eqnarray}
\frac{\text{d}n_{a}}{\text{d}t}&=&-n_{a}(B_{ac}\rho(\omega,v)+\Gamma_{0})\nonumber\\
&&+n_{c}(B_{ca}\rho(\omega,v)+\gamma_{ca})+\mathcal{N}_{a}\Gamma_{0}\\
\frac{\text{d}n_{b}}{\text{d}t}&=&\Gamma_{0}(-n_{b}+\mathcal{N}_{b})+n_{c}\gamma_{cb}\\
\frac{\text{d}n_{c}}{\text{d}t}&=&n_{a}B_{ac}\rho(\omega,v)\nonumber\\
&&-n_{c}(B_{ca}\rho(\omega,v)+\gamma_{c}+\Gamma_{0}),
\end{eqnarray}
where $\mathcal{N}_{a}$ and $\mathcal{N}_{b}$ are the fractional equilibrium populations of states $\vert a\rangle$ and $\vert b\rangle$, respectively, when $B_{ac}=0$.

The equilibrium populations depend upon the thermal distribution of energies \cite{Corney, Demtroder}, but for most atoms the ground state splitting is of the order GHz, while the excited state is separated by hundreds of THz from the ground state. The mean energy per degree of freedom is $\frac{1}{2}k_{B}T$, which at room temperature corresponds to a frequency of THz and so we may assume the ground states to be equally populated and the excited state population to be negligible. The equilibrium population of state $\vert a\rangle$ is therefore 
\begin{equation}
\mathcal{N}_{a}=\frac{g_{a}}{g_{a}+g_{b}}N_{V}.
\end{equation}  
When the rate of the laser scanning across the resonance is much slower than any depopulation mechanisms in the system, we may assume that the populations have reached their steady state values ($\dot{n}_{a}=\dot{n}_{b}=\dot{n}_{c}=0$) and hence become
\begin{eqnarray}
n_{a}&=&\frac{n_{c}\left(B_{ca}\rho(\omega,v)+\gamma_{ca}\right)+\mathcal{N}_{a}\Gamma_{0}}{B_{ac}\rho(\omega,v)+\Gamma_{0}},\label{ssa}\\
n_{c}&=&\frac{n_{a}B_{ac}\rho(\omega,v)}{\gamma_{c}+\Gamma_{0}+B_{ca}\rho(\omega,v)}\label{ssc},
\end{eqnarray}
where we assume $\vert\omega-\omega_{ca}\vert\ll\vert\omega-\omega_{cb}\vert$ and hence may neglect $n_{b}$. This makes the derivation valid for any \emph{open} two-level system. We then substitute Equation \ref{ssc} into \ref{ssa} and rearrange to find $n_{a}$,
\begin{equation}
n_{a}=\frac{\mathcal{N}_{a}}{1+\frac{g_{c}}{g_{a}}\beta_{ac}\alpha_{ac}}
\end{equation}
where
\begin{equation}
\alpha_{ac}=\left(1+\frac{\gamma_{c}+\Gamma_{0}}{B_{ca}\rho(\omega,v)}\right)^{-1}
\end{equation}
is the saturation parameter, and
\begin{equation}
\beta_{ac}=1+\frac{\gamma_{c}-\gamma_{ca}}{\Gamma_{0}}\label{eqn60}
\end{equation} 
is the optical pumping parameter which enhances the saturation of the spectra.
As derived in Section \ref{saturation}, the absorption cross section for the incident beam is proportional to the population difference between the ground and excited states, $B_{ac}n_{a}-B_{ca}n_{c}$ which we may find in terms of $\mathcal{N}_{a}$ by using equation \ref{ssc}
\begin{eqnarray}
B_{ac}n_{a}-B_{ca}n_{c}&=&B_{ac}\mathcal{N}_{a}\left(\frac{1-\alpha_{ac}}{1+\frac{g_{c}}{g_{a}}\beta_{ac}\alpha_{ac}}\right)\nonumber\\
&=&B_{ac}\mathcal{N}_{a}\Delta N \label{diff3level}
\end{eqnarray}
where $\Delta N$ is the population difference caused by the pump beams. This may be rearranged into a form similar to Equation \ref{eqn14}
\begin{equation}
\Delta N=\left(\frac{1-\alpha_{ac}}{1+\frac{g_{c}}{g_{a}}\beta_{ac}\alpha_{ac}}\right)=\left[1+\frac{I_{P}}{2I^{R}_{ac}}L(\omega, -v)\right]^{-1},\label{eqn29}
\end{equation}
in which $I^{R}_{ac}$ is the reduced saturation intensity reported in the literature \cite{Pappas, Demtroder} which is valid for an open two level system:
\begin{equation}
I_{ik}^{R}=I_{ik}\left(\frac{1+\Gamma_{0}/\gamma_{k}}{2+(\gamma_{k}-\gamma_{ki})/\Gamma_{0}}\right).
\end{equation} 
The numerator in the first term of Equation \ref{eqn29} accounts for the steady state population pumped into the excited states and the denominator accounts for the population pumped into the ground states.
\begin{figure}[!ht]
\centering
\includegraphics[width=8cm]{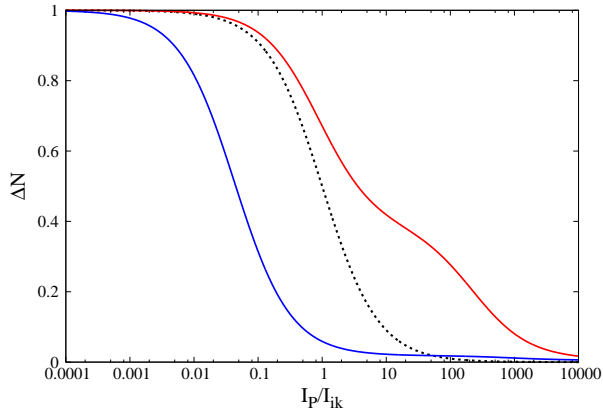}
\caption{\emph{(colour online) The effect of optical pumping in multilevel atoms. The dashed line shows the population difference using Equation \ref{eqn29}, \emph{i.e.}, no optical pumping. The solid blue line shows the population difference for the $^{85}$Rb $F=3\to F'=3$ transition in which optical pumping significantly empties the upper ground state at intensities much lower than saturation. The solid red line shows the population difference for the $^{85}$Rb $F=3\to F'=4$ transition which is normally considered `closed'; we would expect the population to follow the dashed line and equal $1/2$ at the saturation intensity. In the absence of off-resonant pumping the population will tend toward an equilibrium with the transit rate $\Gamma_{0}$ ($\Delta N\sim0.44$ in this instance) and so the population at saturation would equal one half of this value from unity ($1-\frac{1}{2}0.44=0.72$). The equilibrium value is not reached at high intensities due to increased off resonant optical pumping, but evidence for this can be seen in the slight shoulder around $I_{P}/I_{ik}=100$.}}\label{pumping}
\end{figure}
\subsection{Multilevel atoms}\label{multilevel}
Equation \ref{diff3level} may be intuitively extended to include multiple ground ($i$) and excited ($k$) levels.
\begin{eqnarray}
\Delta N(v)&=&\sum_{i}\Delta N_{i}(v)=\sum_{i}\mathcal{N}_{i}\frac{1-\alpha_{ik}(v)}{1+\frac{g_{k}}{g_{i}}\beta_{ik}\alpha_{ik}(v)}\label{eqn30}\\
\alpha_{ik}(v)&=&\sum_{k}\left(1+\frac{\gamma_{j}+\Gamma_{0}}{B^{P}_{ki}\rho(\omega,v)(\omega,v)}\right)^{-1}\\
\beta_{ik}&=&\sum_{k}1+\frac{\gamma_{k}-\gamma_{ki}}{\Gamma_{0}}.
\end{eqnarray}
The absorption of the probe beam is therefore
\begin{equation}
\frac{\text{d}I(\omega)}{\text{d}z}=-\frac{\hbar\gamma}{2\pi c}N_{V}\int_{-\infty}^{+\infty}f_{D}(v)\Delta N(-v) \sum_{i,k}B^{0}_{ik}\rho(\omega,v) \text{d}v.
\end{equation}
Because of optical pumping even at low intensities, the pump beam can no longer be assumed to have a negligible effect upon the spectra. Therefore, the pumping rates in the above model are given by
\begin{eqnarray}
B^{P}_{ki}\rho(\omega,v)=\frac{\gamma^{2}}{8cg_{i}}\Bigg( \frac{I_{P}}{I_{ik}}\frac{\gamma/2}{(\omega-\omega_{ik}-kv)^{2}+(\frac{\gamma}{2})^{2}}\nonumber
\\+\frac{I_{0}}{I_{ik}}\frac{\gamma/2}{(\omega-\omega_{ik}+kv)^{2}+(\frac{\gamma}{2})^{2}}\Bigg)\label{eqn36}
\end{eqnarray}
and
\begin{equation}
B^{0}_{ik}\rho(\omega,v)=\frac{\gamma^{2}}{8cg_{i}}\Bigg(\frac{I_{0}}{I_{ik}}\frac{\gamma/2}{(\omega-\omega_{ik}+kv)^{2}+(\frac{\gamma}{2})^{2}}\Bigg).
\end{equation}
Note that for Equation \ref{eqn36} the Doppler shift for the probe beam is opposite to that of the pump beam.\par
This model may be simplified by considering the steady state population for open or closed transitions with fast (MHz) decay rates. Off-resonant pumping into the dark state must be included both to model crossover peaks correctly \cite{Demtroder} and because an equilibrium is set up between pumping and transit of atoms into the beam (see Figure \ref{pumping}). For an open transition the atoms will be pumped into the dark state before a significant population builds up in the excited state, so one may use the following form
\begin{equation}
\Delta N=\left(1+\sum_{k}\frac{g_{k}}{g_{i}}\alpha_{ik}\beta_{ik}\right)^{-1}.
\end{equation}
For a closed state (such as $^{85}$Rb $F=3\to F'=4$) a significant population will build up in the excited state, but the off-resonant excited state populations will be negligible. In such a situation off-resonant optical pumping into the dark ground states reduces the equilibrium population as shown in Figure \ref{pumping} and Equation \ref{eqn30} may then be simplified to the form
\begin{equation}
\Delta N=\frac{1-\alpha_{ik}}{1+\sum_{k}\frac{g_{k}}{g_{i}}\alpha_{ik}\beta_{ik}}.
\end{equation}

\subsection{Transit broadening}\label{transit}
Transit broadening is normally considered in the context of a limit to the interaction time and hence to the fundamental resolution of the spectra. For a Gaussian laser beam with a $1/e^{2}$ radius $R$ passing through a dilute gas of atoms with a mean velocity $u$, the standard result \cite{Shimoda, Demtroder} for the transit rate of an atom through the beam is
\begin{equation}
\Gamma_{0}=\frac{2u\sqrt{\ln 2}}{R}.
\end{equation}
However, we can see in Equation \ref{eqn60} that the transit rate has a significant effect on the optical pumping rate. It is common practice to use large beams to reduce the transit broadening but this has the effect of increasing the time during which atoms may be optically pumped into the dark state, and hence acts to \emph{decrease} the resolution. In Figure \ref{pumping} we see that optical pumping occurs at intensities much lower than $I_{ik}$; in fact, the atom can be pumped into the dark state $\vert b\rangle$ with the absorption of a single photon and so this complicates the calculation of the beam width. If we assume that significant optical pumping occurs at a specific intensity, $I_{\varphi}$, and calculate the width of beam at this intensity as the total pump intensity is changed, the transit rate becomes;
\begin{equation}
\Gamma_{ik}=\left\{\begin{array}{cr}
\Gamma_{0}\left[\sqrt{\frac{1}{2}\ln\left(\frac{I_{P}}{I_{\varphi}}\right)}\right]^{-1}\qquad &\text{if } I_{P}>I_{\varphi} \\
\Gamma_{0} \qquad &\text{if } I_{P}\leq I_{\varphi} \label{eqn40}
\end{array}\right.
\end{equation}
The transit rate here depends upon the specific transition ($i\to k$), since the apparent increase of beam width will depend upon the individual transition strength. It is not obvious what value of $I_{\varphi}$ of should be used; however in the following model we find empirically that the reduced saturation intensity ($I_{\varphi}=I^{R}_{ik}$) fits the data well.
\subsection{Additional broadening mechanisms}\label{broaden}
The experimental setup may introduce extra broadening of the spectral features and although they are usually much less than the natural and Doppler linewidths, they can be on the order of the transit rate.

The following broadening mechanisms have a Lorentzian lineshape or effect the frequency detuning, therefore may be included to our model by summation with the natural decay rate \cite{Corney}.
\subsubsection{Laser linewidth}
As mentioned earlier, the resolution of the spectral features is constrained by the tool we use to measure them: the laser. Diode lasers are a common tool for pump probe spectroscopy and may have linewidths down to the hundreds of kHz region with the addition of an external cavity. We assume a Lorentzian spectral laser line-shape with a full width at half maximum (FWHM) $\Gamma_{L}$ in angular frequency units.  
\subsubsection{Geometrical broadening}
The sub-Doppler resolution depends upon atoms traversing the beams at a perpendicular angle. Therefore, if the beams are not exactly anti-parallel they will sample a non-zero atomic velocity component and hence decrease the resolution. In the experimental arrangement described in Section \ref{apparatus}, perfect beam overlap is achieved using a polarizing beam--splitting cube. For some experiments the polarizations of each beam may need to be controlled independently, and back-reflections into the laser cavity may be unwanted, favoring a geometry in which the counter-propagating beams cross at a non-zero angle $\theta$. This results in a broadening \cite{Shimoda}
\begin{equation}  
\Gamma_{G}=(\vec{k}_{P}-\vec{k}_{0})\cdot \vec{v}=\frac{2u\omega_{ik}}{c}\,\sin(\tfrac{1}{2}\theta).
\end{equation}
where $\vec{k}_{P}$ and $\vec{k}_{0}$ are the pump and probe wavevectors, respectively.
\subsubsection{Beam collimation}
An uncollimated beam passing through the vapor cell will result in a variation of transit time along the cell. Wavefront curvature also broadens the Lamb dip in much the same manner as geometrical broadening. To limit this effect the radius of curvature of the wavefronts must be much greater than $kR^{2}/4$ \cite{Letokhov}.
\subsubsection{Collisional broadening}
We have limited our model to dilute gases in which collisions during the interaction time are negligible. In dense gases the collisional cross-section can be as large as the absorption cross-section and can affect the spectra by broadening and shifting the peak. Collisions may also affect the distribution of population amongst the ground states. The magnitude and nature of the effect depend upon density, temperature and collisional partners (\emph{i.e.}, non-identical atoms in the case of a buffer gas) \cite{Corney}.
 
\section{Rubidium}\label{rubidium}
\begin{figure}
\centering
\includegraphics[width=8cm]{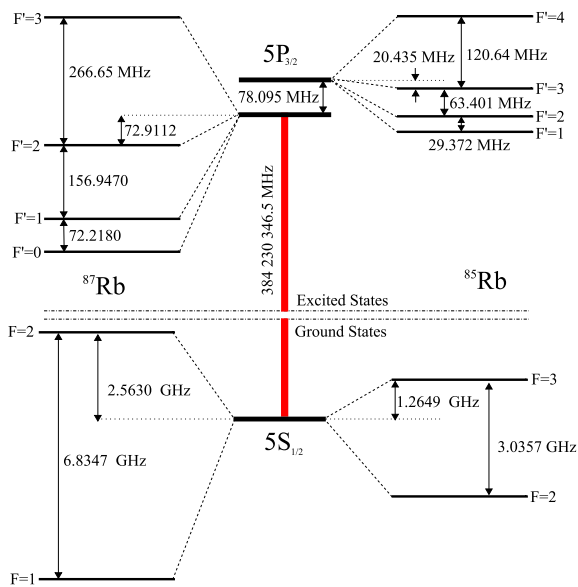}
\caption{\emph{Hyperfine structure of $^{85}$Rb and $^{87}$Rb.}\label{freqs}}
\end{figure}

Natural rubidium occurs in two stable isotopes, $^{85}$Rb and $^{87}$Rb, with fractional abundances, $f_{I}$, of 0.2783 and 0.7217, and atomic masses 84.912 and 86.909, respectively \cite{Lide}. The ground state is $5S_{1/2}$, with nuclear moment $I=5/2$ for $^{85}$Rb and $I=3/2$ for $^{87}$Rb, and orbital angular momentum $L=0$ and spin $S=1/2$. The excited state under investigation is $5P_{3/2}$ state with $L=1$ and the transition between these states is known as the $D_{2}$ line. Due to the non-zero nuclear angular momentum each state is split into $2J+1$ hyperfine levels, each of which has $2(2I+1)$ magnetic sublevels. Figure \ref{freqs} shows the hyperfine structure of both isotopes and the energy splitting between levels. The Clebsch-Gordan coefficients, saturation intensity (Equation \ref{eqn15}) and fractional decay rates $\gamma_{ik}$ used in the model are shown in Table \ref{Rbvalues}. The natural lifetime is $26.25\pm0.07\thinspace$ns \cite{Schultz} for both isotopes. 

\begin{table}[!t]
{\renewcommand{\arraystretch}{1.2}
\renewcommand{\tabcolsep}{0.2cm}
\begin{tabular}{c||c|c|c|c|c}
{}&$F_{i}$&$F_{k}$&$I_{ik}$(mW\,cm$^{-2}$)&$C^{2}_{ik}$&$\gamma_{ki}(\gamma)$\\
\hline
\hline
$^{85}$Rb&3&4&{3.894}&1&1\\
{}&3&3&{9.012}&35/81&$5/9$\\
{}&3&2&{31.542}&10/81&$2/9$\\
{}&2&3&{8.046}&28/81&$4/9$\\
{}&2&2&{6.437}&35/81&$7/9$\\
{}&2&1&{8.344}&1/3&$$\\
\hline
$^{87}$Rb&2&3&3.576&7/9&$1$\\
{}&2&2&10.013&5/18&$1/2$\\
{}&2&1&50.067&1/18&$1/6$\\
{}&1&2&6.008&5/18&$1/2$\\
{}&1&1&6.008&5/18&$5/6$\\
{}&1&0&15.020&1/9&$1$\\
\end{tabular}}
\caption{Rubidium model parameters for linear polarization and degenerate hyperfine levels.}\label{Rbvalues}
\end{table}
Rubidium has a melting point of 39.3\textcelsius and the number density is then given by \cite{Nesmeyanov}
\begin{equation}
N_{V}=\frac{f_{I}}{k_{B}T}133.323\times10^{p_{j}}\\
\end{equation}
where the exponent subscript $j=S,L$ corresponds to solid and liquid rubidium, respectively, with
\begin{eqnarray}
p_{S}&=&-94.04826-\frac{1961.258}{T}\nonumber\\
&&-0.03771678\times T+42.57526\ln(T)\\
p_{L}&=&15.88253-\frac{4529.535}{T}\nonumber\\
&&+0.00058663\times T-2.99138\ln(T).
\end{eqnarray}
 
\section{Pump--Probe Apparatus}\label{apparatus}
We have tested our model experimentally with rubidium vapor in a pump-probe
apparatus, normally used to stabilize diode lasers for a magneto-optical
trap, shown in Figure 4. The diode laser had an external cavity in the
Littrow configuration with a beam diameter of $2.1 \pm 0.1$\,mm in the
vertical plane. The beam was elliptical with a horizontal width equal to
twice the height; the greatest transit broadening thus results from the
narrower dimension and the model therefore uses the above value. The beams
were overlapped through the vapor cell by sending the pump beam through a
polarizing beam splitter cube (PBSC); the counterpropagating probe beam had
a linear polarization perpendicular to the pump beam and was reflected by
the PBSC onto the photodiode. This layout allowed perfect overlap of the
beams and a reduced footprint of the apparatus.

The laser frequency was
scanned via rotation of the external cavity grating with a piezoelectric
transducer and spectra were averaged over three scans. The scans were not
linear, so the frequency axis was calibrated using the 12 resonance and
crossover peaks (see Figure \ref{freqs}) by fitting to a fourth-order polynomial; the standard
deviation of the fitted peaks from tabulated values was $<400$\,kHz. The probe
power was kept at $1 \mu$W and the laser linewidth was $1.0 \pm 0.2$\,MHz
[28]. The vapor cell had  length of 75~mm and a temperature of
$18.0\pm0.1$\,\textcelsius. No attempt was made to null the Earth's or nearby magnetic
fields. 
\begin{figure}[!t]
\centering
\includegraphics[width=7cm]{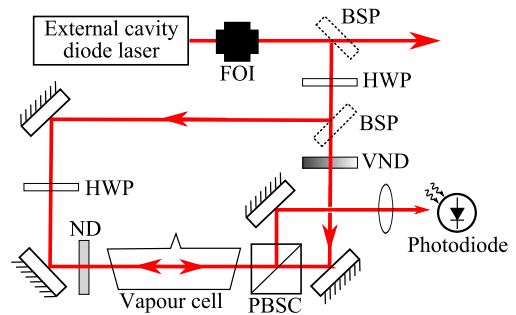}
\caption{\emph{The pump--probe spectroscopy apparatus used for the spectra in Section \ref{compare}. Approximately 3\,mW is picked off from the external cavity diode laser (ECDL) with a beam--splitter plate (BSP). The ECDL is protected from back reflection by a Faraday Optical Isolator (FOI), so that the beam may be perfectly overlapped using a polarizing beam splitter cube (PBSC). The linear polarization of the probe (left to right through the vapor cell) and pump (right to left) beams are rotated by half waveplates (HWP) so that the pump is transmitted and the probe reflected by the PBSC. The power of the pump beam is controlled with a variable neutral density filter (VND). }\label{layout}}
\end{figure}
\section{Comparison between Experiment and Theory}\label{compare}
The experimentally measured pump probe spectra for the upper ground states of $^{85}$Rb and $^{87}$Rb are shown together with predictions of the theoretical model in Figure \ref{spectra}$i$, where $i$=$a$,$b$,$c$,$d$ relate to the pump powers 20$\,\mu$W, 100$\,\mu$W, 500$\,\mu$W and 2500$\,\mu$W, respectively. Figure \ref{spectra} is split into four parts: plots $i$-3 shows the full absorption spectrum with the experimental data in black and the theoretical curves in red, plots $i$-1,2 show a magnified sections of the Lamb dips ($i$-1=$^{87}$Rb, $i$-2=$^{85}$Rb) and plots $i$-4 shows the experimental data subtracted from the theoretical curves (residuals).

\begin{figure}
\centering
\includegraphics[width=7cm]{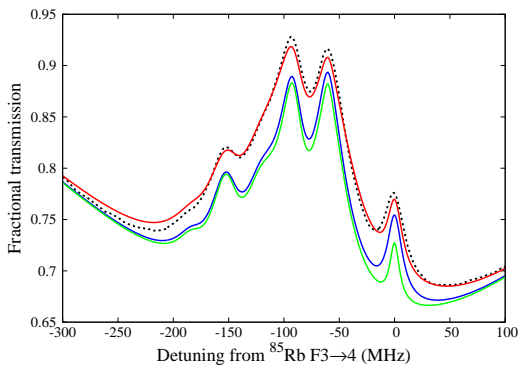}
\caption{\emph{(colour online) The effect of beam width on sub-Doppler features. The dashed line and solid red line correspond to the Figure \ref{spectra}d-2. The blue line shows the same model but without the additional beam broadening factor from Equation \ref{eqn40}. The green line includes the additional broadening factor but assumes a $1/e^{2}$ beam width twice that in the red plot. The laser power remains the same for each plot.}\label{beamwidth}}
\end{figure}
\begin{figure*}[t]
\centering
\includegraphics[height=18cm]{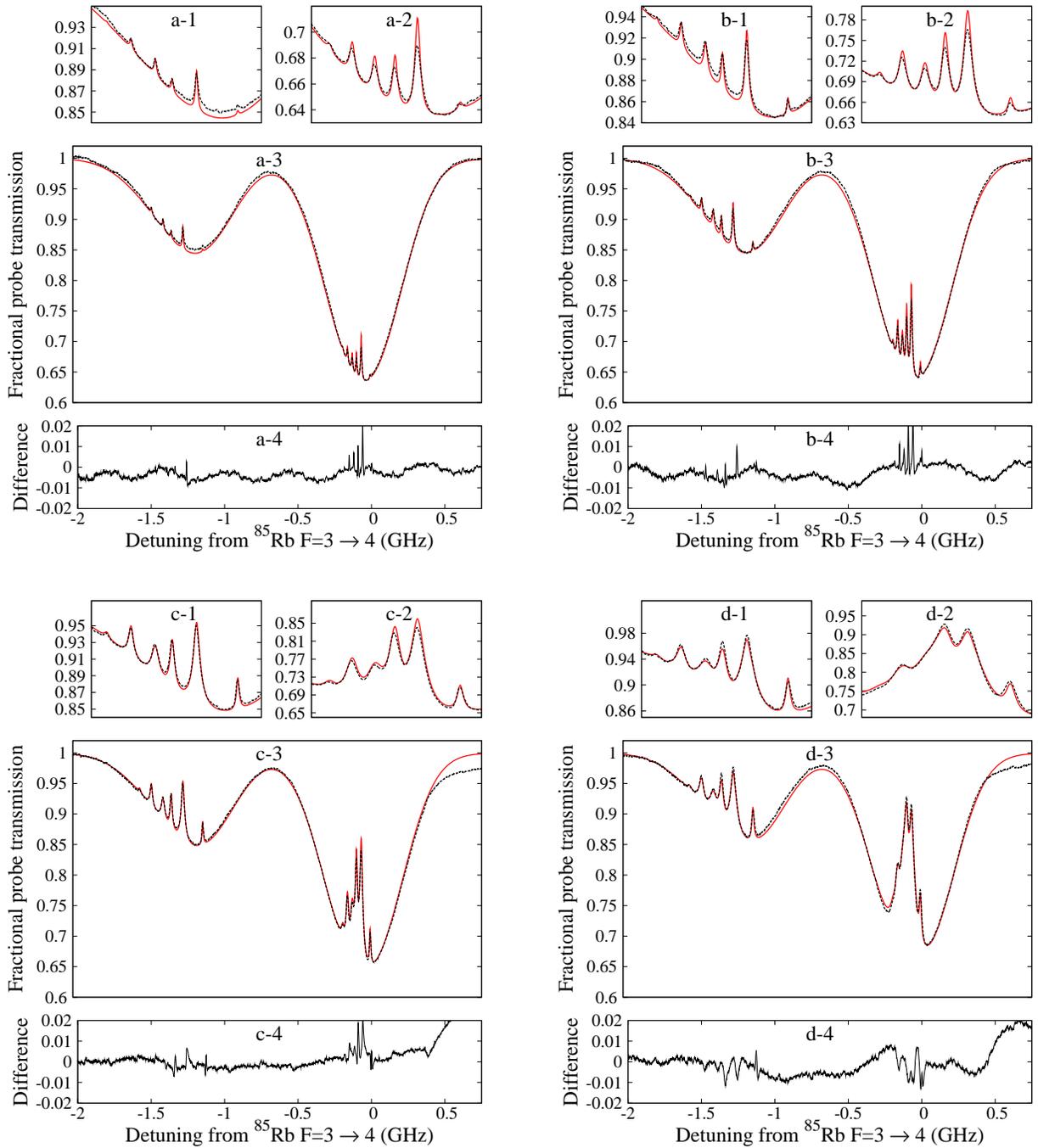}
\caption{\emph{Normalized pump probe transmission spectra of $^{85}$Rb $F=3\to4$ (right hand dip) and $^{87}$Rb $F=2\to3$ (left-hand dip) for pump powers of 20$\,\mu$W, 100$\,\mu$W, 500$\,\mu$W and 2500$\,\mu$W, relating to a,b,c,d respectively. The experimental data are shown by black dashes and the theoretical curves by the solid red line. Plots i-1,2 are magnified sections of the full absorption spectra i-3, and i-4 is the difference between the data and model. The large deviations in Figures c,d-3,4 above +0.5\,GHz are due to the nonlinear scanning of the piezo near a turning point.}\label{spectra}}
\end{figure*} 
It is apparent that our theoretical model accurately predicts the height and width of each absorption line. The residuals have a standard deviation of less than 1\%, and much of this is due to experimental noise and calibration errors. In Figures 6a-4 and 6b-4, the
noise is dominated by a sinusoidal signal from power line pickup.

Our model appears to slightly overestimate the height of the Lamb dips at low powers. This may be due to errors in the beam power measurement or the elliptical cross section of the laser beams. This latter point complicates the effect of beam width on optical pumping, and the arbitrary value used for $I_{\varphi}$ in Section \ref{transit}. The effect of these values is shown in Figure \ref{beamwidth}. 

\section{Conclusion}\label{conc}
A simple model of pump--probe spectroscopy based on the rate equations has been presented and compared against the experimental spectrum of rubidium. Our model can describe multilevel atoms and fits the data well with the residual difference less than 1\% over a large range of pump powers. The model is valid for any dilute gas when coherent effects are negligible, and accounts for finite laser linewidth, optical pumping and transit time broadening.

The most significant effect upon the spectral features is the optical pumping during the atom's transit across the beam. As opposed to saturation broadening, which is due to the equalization of populations in the ground and excited states preventing further absorption, optical pumping can transfer population into dark states, which also prevents further absorption but may occur for a single photon absorption. The time taken by the atom to transverse the beam significantly affects the optical pumping and therefore careful attention must be made in defining the beam width.
\begin{acknowledgments}
The authors would like to thank Dr Ifan Hughes for stimulating discussions in the course of this work. 
\end{acknowledgments}

\end{document}